\documentstyle[12pt,fleqn]{article}
\def\nel{N_{\rm el}}

\def\beeq{\begin{equation}}
\def\eneq{\end{equation}}
\def\beeqa{\begin{eqnarray}}
\def\eneqa{\end{eqnarray}}

\addtocounter{section}{-1}
\begin{document}

\begin{center}

\vspace{2in}

{\large {\bf{Exciton effects in soliton and bipolaron lattice states\\
of doped electron-phonon Peierls systems
} } }

\vspace{1cm}

(Running head: {\sl Exciton effects in doped Peierls systems})

\vspace{1cm}

{\rm Kikuo Harigaya\footnote[1]{Electronic mail address:
harigaya@etl.go.jp.}, Yukihiro Shimoi, and Shuji Abe
}\\

\vspace{1cm}

{\sl Fundamental Physics Section,\\
Electrotechnical Laboratory,\\
Umezono 1-1-4, Tsukuba, Ibaraki 305, Japan}

\vspace{1cm}

(Received~~~~~~~~~~~~~~~~~~~~~~~~~~~~~~~~~~~)
\end{center}

\Roman{table}

\vspace{1cm}

\noindent
{\bf Abstract}\\
Exciton effects on soliton and bipolaron lattice states are
investigated using an electron-lattice Peierls model with
long-range Coulomb interactions.  The Hartree-Fock (HF)
approximation and the single-excitation configuration-interaction
(single-CI) method are used to obtain optical absorption
spectra.  We discuss the following properties:
(1)  The attraction between the excited electron and the
remaining hole makes the excitation energy smaller when
the correlations are taken into account by the single-CI.
The oscillator strengths of the lower excited states become
relatively larger than in the HF calculations.
(2)  We look at variations of relative oscillator strengths
of two or three kinds of excitons described by the single-CI.
While the excess-electron concentration is small, the ratio
of the oscillator strengths of the exciton with the lowest
energy increases almost linearly.  The oscillator strengths
accumulate at this exciton as the concentration increases.

\mbox{}


\pagebreak

\section{Introduction}

It has been well known that correlation effects are present
among $\pi$-electrons in conjugated polymers.  For example:
\begin{itemize}
\item The envelope of the wavefunction of the midgap state of
the neutral (spin) soliton in {\sl trans}-polyacetylene has
sites with spin density of the opposite sign [1].  This fact
of the ``negative spin density" has been explained by considering
Coulomb interactions in the Su-Schrieffer-Heeger (SSH) model
[2] of conjugated polymers [3].
\item The electron paramagnetic resonance experiment of
pernigraniline base [4], reported recently, has studied the
spin distribution around a neutral soliton.  It has been
suggested that sites with negative spin density exist as
the consequence of correlation effects.
\item Nonlinear optical response functions of polydiacetylenes
exhibit excitation structures owing to the presence of excitons.
Their structures have been explained theoretically by using the
intermediate exciton formalism [5].
\end{itemize}
Therefore, it is generally interesting to study Coulomb
interaction effects in conjugated polymers.  In fact, this
problem has been considered by various authors for more than
a decade since the SSH model [2] was proposed.

Electronic excitation structures in the half-filled conjugated
polymers with the constant dimerization have been theoretically
investigated by using the exciton formalism [5] and the exact
diagonalization method [6], and also by solving the time-dependent
Hartree-Fock (HF) equations [7].  This problem was pointed out
earlier, but excitation structures have been considered intensively
only recently, relating with origins of nonlinear optical
spectra [6-8].  The most remarkable consequence of correlation
effects is that the lowest energy excitation has the largest
oscillator strength.  It is clearly seen when the optical spectra
calculated by using the HF wavefunctions are compared with the
spectra with the correlation effects.  This fact does not depend
on whether the higher correlations are taken into account by the
single-excitation configuration-interaction (single-CI) method [5],
or by the time-dependent HF formalism [7].

It is widely known that the soliton, polaron, and bipolaron
lattices are present [9], when the SSH model [2], its continuum
version, and the extended model with the term of the nondegeneracy
[10] are doped with electrons or holes.  New bands related with
nonlinear excitations develop in the Peierls gap as the doping
proceeds.  When correlation effects are considered by the single-CI,
the excitation structures exhibit the presence of excitons.
There is one kind of exciton in the half-filled system, where the
excited electron (hole) sits at the bottom of the conduction
band (top of the valence band).  We will call this as the
``intercontinuum exciton".  In the soliton lattice states of the
doped SSH model for degenerate conjugated polymers, there are
small gaps between the soliton band and the continuum states,
i.e., valence and conduction bands.  Therefore, the number of the
kind of excitons would increase and their presence will be
reflected in structures of the optical spectra.  A new exciton,
namely, ``soliton-continuum exciton" will appear when the
electron-hole excitation is considered between the soliton and
one of the continuum bands.  The main purpose of this paper is
to embody the above picture.  There have been a lot of investigations
of correlation effects in doped SSH systems, but the work in the
view point of excitons has been rarely performed.  General properties
of exciton effects on soliton lattice systems are the first
interest of the paper.  The portion of the spectral weights
of the two kinds of excitons will be calculated and discussed.

There are two midgap bands, when the nondegenerate conjugated
polymers are doped and bipolaron lattice states are formed.
The excitation spectra will become complicated due to the increase
of number of kinds of excitons.  The second part of this paper will
be devoted to this problem, and the relative spectral weight
of each exciton will be studied again.

This paper is composed as follows.  In \S 2, the model is
introduced and the numerical method is explained.  Results for
the soliton lattice system are reported in \S 3.  Effects of the
nondegeneracy are investigated in \S 4.  The paper is summarized
in the final section.

\section{Model}

The following hamiltonian is used to discuss excitonic effects
in soliton and bipolaron lattice states of Peierls systems:
\beeq
H = H_{\rm SSH} + H_{\rm int}.
\eneq
The first term $H_{\rm SSH}$ of Eq. (1) is the SSH model:
\beeqa
H_{\rm SSH} &=& - \sum_{i,\sigma} [t - \alpha y_i + (-1)^i \delta_0 t]
( c_{i,\sigma}^\dagger c_{i+1,\sigma} + {\rm h.c.} ) \\ \nonumber
&+& \frac{K}{2} \sum_i y_i^2,
\eneqa
where $t$ is the hopping integral of the system without the
dimerization; $\alpha$ is the electron-phonon coupling constant
which changes the hopping integral linearly with respect to the
bond variable $y_i$; $\delta_0 t$ is the Brazovskii-Kirova (BK) term
which measures the degree of the nondegeneracy (it was originally
introduced in the continuum model [10]); $c_{i,\sigma}$ is an
annihilation operator of the $\pi$-electron at the site $i$ with
spin $\sigma$; the sum is taken over all the lattice sites of the
periodic chain; and the last term with the spring constant $K$
is the harmonic energy of the classical spring simulating the
$\sigma$-bond effects.  The second term of Eq. (1) is the
long-range Coulomb interaction in the form of the Ohno potential:
\beeqa
H_{\rm int} &=& U \sum_i
(c_{i,\uparrow}^\dagger c_{i,\uparrow} - \frac{n_{\rm el}}{2})
(c_{i,\downarrow}^\dagger c_{i,\downarrow} - \frac{n_{\rm el}}{2})\\ \nonumber
&+& \sum_{i \neq j} W(r_{i,j})
(\sum_\sigma c_{i,\sigma}^\dagger c_{i,\sigma} - n_{\rm el})
(\sum_\tau c_{j,\tau}^\dagger c_{j,\tau} - n_{\rm el}),
\eneqa
where $n_{\rm el}$ is the number of $\pi$-electrons per site,
$r_{i,j}$ is the distance between the $i$th and $j$th sites, and
\beeq
W(r) = \frac{1}{\sqrt{(1/U)^2 + (r/r_0 V)^2}}
\eneq
is the Ohno potential.  The quantity $W(0) = U$ is the strength of
the onsite interaction, and $V$ means the strength of the long range part.

The model is treated by the HF approximation and the single-CI
for the Coulomb potential.  The adiabatic approximation is forced
on the bond variables.  The HF order parameters and bond variables
are determined selfconsistently using the standard iteration
method [11].  After the HF approximation $H \Rightarrow H_{\rm HF}$,
we divide the total hamiltonian as $H = H_{\rm HF} + H'$.  The term
$H'$ becomes
\beeqa
H' &=& U \sum_i
(c_{i,\uparrow}^\dagger c_{i,\uparrow} - \rho_{i,\uparrow})
(c_{i,\downarrow}^\dagger c_{i,\downarrow} - \rho_{i,\downarrow})\\ \nonumber
&+& \sum_{(i,j),i \neq j} W(r_{i,j})
[\sum_{\sigma,\tau} ( c_{i,\sigma}^\dagger c_{i,\sigma} - \rho_{i,\sigma})
(c_{j,\tau}^\dagger c_{j,\tau} - \rho_{j,\tau})\\ \nonumber
&+& \sum_\sigma ( \tau_{i,j,\sigma} c_{j,\sigma}^\dagger c_{i,\sigma}
+ \tau_{j,i,\sigma} c_{i,\sigma}^\dagger c_{j,\sigma}
- \tau_{i,j,\sigma} \tau_{j,i,\sigma})],
\eneqa
where $\rho_{i,\sigma} = \langle c_{i,\sigma}^\dagger c_{i,\sigma} \rangle$
and $\tau_{i,j,\sigma} = \langle c_{i,\sigma}^\dagger c_{j,\sigma} \rangle$
are Hartree-Fock order parameters.  When we write the Hartree-Fock ground
state  $| g \rangle = \prod_{\lambda {\rm: occupied}}
c_{\lambda,\uparrow}^\dagger c_{\lambda,\downarrow}^\dagger
|0 \rangle$ and the single electron-hole excitations
$|\mu \lambda \rangle = c_{\mu, \sigma}^\dagger
c_{\lambda, \tau} | g \rangle$ ($\mu$ means an unoccupied state;
we assume both singlet and triplet excitations in this abbreviated
notation), the matrix elements of the HF part and the excitation
hamiltonian become as follows:
\beeqa
\langle \mu' \lambda' | ( H_{\rm HF} - \langle H_{\rm HF} \rangle)
| \mu \lambda \rangle &=& \delta_{\mu',\mu} \delta_{\lambda',\lambda}
(E_\mu - E_\lambda),\\
\langle \mu' \lambda' | ( H' - \langle H' \rangle) | \mu \lambda \rangle
&=& 2J \delta_S - K,
\eneqa
where $E_\mu$ is the energy of the HF orbital,
$\delta_S = 1$ for spin singlet, $\delta_S = 0$ for spin triplet, and
\beeqa
J(\mu',\lambda';\mu,\lambda) &=& \sum_{i,j} V_{i,j}
\langle \mu' | i \rangle \langle \lambda' | i \rangle
\langle j | \mu \rangle \langle j | \lambda \rangle \\
K(\mu',\lambda';\mu,\lambda) &=& \sum_{i,j} V_{i,j}
\langle \mu' | j \rangle \langle \lambda' | i \rangle
\langle j | \mu \rangle \langle i | \lambda \rangle
\eneqa
with $V_{i,i} = U$, $V_{i,j} = W(r_{i,j})$ for $i \neq j$.  The
diagonalization of the total hamiltonian $H$ gives the set of the
excited states $\{ | \kappa \rangle \}$ within the single-CI method.
In the actual calculation, we limit the spin configurations to the
singlet excitations which are the main interests of optical excitations.

We assume a geometry of a ring for a polymer chain, in order to
remove edge effects.  If we use an open boundary, the dimerization
and thus the Peierls gap becomes larger near the two edges, and
this might result in artifacts of optical spectra.  We shall use
the coordinate of $j$th carbon atoms,
\beeq
(r \cos \frac{2 \pi j}{N}, r \sin \frac{2 \pi j}{N}, 0),
\eneq
where $r = Na/(2\pi)$ is the radius of the polymer ring; $N$
is the system size and $a$ is the lattice constant.  The electric
field of light is parallel to the $x$-$y$ plane.  In order to
obtain optical spectra which are independent of the relative
positions of solitons with respect to the direction of light,
we shall sum up two spectra where light is along with the $x$-
and $y$-directions.  So, we use the following formula of the spectrum:
\beeq
\sum_\kappa E_{\kappa} P (\omega - E_{\kappa})
(\langle g | x |\kappa \rangle\langle \kappa | x | g \rangle
+ \langle g | y |\kappa \rangle\langle \kappa | y | g \rangle).
\eneq
Here, $P (\omega) = \gamma/[ \pi (\omega^2 + \gamma^2)]$
is the Lorentzian distribution ($\gamma$ is the width),
$E_{\kappa}$ is the electron-hole excitation energy, and
$| g \rangle$ means the ground state.  In eq. (11),
the quantity,
\beeq
E_{\kappa} (\langle g | x |\kappa \rangle\langle \kappa | x | g \rangle
+ \langle g | y |\kappa \rangle\langle \kappa | y | g \rangle),
\eneq
is the oscillator strength of the excited state $| \kappa \rangle$.
Applying electric field to the ring-shaped polymer simulates the
situation that polymer chains are oriented randomly in every direction
within the $x$-$y$ plane.  The average over orientations are
effectively performed.  The similar idea has been used in the
recent paper by Abe et al [5].

The system size is chosen as $N= 80, 100, 120$ when the electron
number is even (it is varied from $\nel = N, N+2, N+4, N+6$ to $N+8$),
because the size around 100 is known to give well the energy gap
value of the infinite chain.  More larger system size becomes
tedious for doing single-CI calculations which call for huge computer
memories.  When there is one soliton, we take $N = 81, 101, 121$,
and use the periodic boundary condition also.

In principle, we have to adjust parameters and find appropriate
ones in order to reproduce experimental data, such as, the energy
gap and the dimerization amplitude.  But, we will change parameters
arbitrary in a reasonable range in order to look at excitonic
effects clearly.  The Coulomb parameters are changed within
$0 \leq V \leq U \leq 5t$, and we show results for $U = 2V =2t$
and $=4t$ as the representative cases.  Other parameters, $t = 1.8$eV,
$K = 21$eV/\AA$^2$, and $\alpha = 4.1$eV/\AA, are fixed in view
of the general interests of this paper.  All the quantities of
energy dimension are shown in the units of $t$.

\section{Soliton lattice systems}

Figure 1 shows the typical lattice configuration and excess-electron
density distribution for $N=100, \nel = 104, U=4t, V=2t$, and
$\delta_0 = 0$.  Both quantities are the smoothed data after the
removal of small oscillations between even- and odd-number sites.
There are four charged solitons due to the excess-electron number
$\nel - N =4$.  Solitons are arrayed equidistantly.  The excess-electron
density shows the oscillation of the charge density with
its maxima at the soliton centers.

Next, let us look at optical spectra and consider exciton effects.
We calculate also the optical spectra from single-electron
excitations among HF energy levels as references of exciton
effects.  The typical optical spectra within the HF approximation
and with the single-CI are shown as Figs. 2 and 3, respectively.
The broadening $\gamma = 0.05t$ is used.  The Coulomb parameters
are $U=4t$ and $V=2t$.  Relatively strong Coulomb interactions
are taken in order to look at the exciton effects in the
optical response clearly.  The system size and the electron number
are $(N,\nel) = (101,102), (100,102), (100, 104)$, for
(a), (b), and (c) (of Figs. 2 and 3), respectively.

In order to characterize properties of optical excitations,
in other words, to identify whether the excitons are formed between
valence and conduction bands (``intercontinuum exciton"),
or between soliton and conduction bands (``soliton-continuum
exciton"), we calculate the component of the electron-hole
excitation between the soliton and conduction bands for the
absorption spectra within HF as well as those with the single-CI.
Results are shown in Figs. 2 and 3, superposed with the optical
spectra.  We can easily determine the position of the optical gap
by comparing components of each kind of exciton.  The optical
gap of one exciton is located at the lowest energy where the
component is larger than that of the other exciton.

As we proceed from Fig. 2(a) to Fig. 2(c) with increasing the
soliton concentration, the contribution from the transition
between the soliton band and the valence band becomes larger
than that between the soliton and conduction bands.  In other
words, the optical transition between the soliton and conduction
bands rapidly develops as the soliton concentration increases.
The energy positions of optical gaps of excitons are shown by
the triangles at the top of each figure.  The lowest optical
gap is about 0.7$t$, 0.9$t$, and 1.0$t$ in Figs. 2(a), (b),
and (c), respectively, and is slightly increasing.  The almost
constant behavior can be explained as follows.  The HF order
parameter $\tau_{i,j,\sigma}$ introduces an additional bond
order, and thus increases the energy gap.  In contrast, the lowest
optical gap is a decreasing function of the soliton concentration
in the free-electron case.  Therefore, the decrease of the optical
gap in the free electron case is suppressed by the increase of
the energy gap in the presence of long-range Coulomb interactions.
The optical gap of the transition between the continuum states
is about 1.6$t$, 1.7$t$, and 1.9$t$ for the three figures.
This quantity becomes larger with the concentration, due to
the increase of number of states in the soliton band.

We shall look at the optical spectra by HF and single-CI
calculation in order to discuss exciton effects.  They are shown
in Fig. 3.  The optical gap of the soliton-continuum exciton is 0.6$t$,
0.7$t$, and 0.8$t$ in Figs. 3(a), (b), and (c), respectively.
The optical gap of the intercontinuum exciton is 1.4$t$, 1.4$t$,
and 1.6$t$ for the three figures.  Here, we have regarded that
the energy position of the lowest excitation where the
component of the soliton-continuum exciton becomes smaller
than 0.5, is the optical gap of the intercontinuum exciton.
Both optical gaps decrease apparently from those of Fig. 2.
This is due to the binding of an electron and a hole in the CI
treatment.  We also find that the soliton-continuum exciton
has the larger total oscillator strength than that of Fig. 2.
This is due to the one dimensionality, discussed in ref. [5].

There are many small structures in the optical spectra due
to the finite system size.  They could be removed by doing
calculations for larger systems, as has been done in the
paper [12] with reducing the dimension of the matrix of CI
excitations by means of the translational symmetry for the
half-filled hamiltonian.  But, the reduction of the system
size is difficult for soliton lattice states owing to the
periodicity of the system which changes with the soliton
concentration.  In order to pursue the change of optical
excitation characters systematically for various combinations
of parameters, we rather perform calculations for small
system sizes around 100 and analyze numerical data as
functions of the soliton concentration.  In fact, when
the optical gap and the ratio of the total oscillator
strengths of the soliton-continuum exciton are plotted
against the soliton concentration, $(\nel - N)/N$, the
plots are arrayed rather smoothly.  We shall look at the
data in Figs. 4 and 5.

Figure 4 summarizes the optical gaps of the two kinds of
excitons.  They are calculated by the HF followed by the
single-CI.  Figures 4(a) and (b) are for $U = 2V = 2t$
and $=4t$, respectively.  The gaps of the intercontinuum
exciton and soliton-continuum exciton are shown by closed
and open squares, respectively.  The optical gap of the
soliton-continuum exciton is almost independent of the
concentration, owing to the balance between the decreasing
due to the soliton concentration change and the widening
of the energy gap owing to the presence of long-range
Coulomb interactions.  The gap of the intercontinuum exciton
increases rapidly when the concentration is larger than 2.5
percent.  This is due to the fact that the number of states
in the soliton band increases, and thus the energy gap
between continuum states increases.  It seems that these
properties are common for the two Coulomb parameter sets.
The optical gaps are larger for stronger Coulomb interactions.
This is due to the larger bond order parameters which enhance
the Peierls gap of the system.

Figure 5 shows the ratio of the oscillator strengths of
the soliton-continuum exciton, plotted against the soliton
concentration.  The closed and open squares are the results
of the HF-CI and HF calculations, respectively.  The closed
squares have the larger ratio than the open ones.  This is
one of exciton effects.  When the concentration is near zero,
the ratio varies almost linearly.  This would be the natural
consequence for low concentrations, because interactions
among solitons are exponentially small and thus the ratio is
proportional to the number of solitons.  The increase near
the zero concentration is slower for the stronger $U$
and $V$ of Fig. 5(b) than in Fig. 5(a).  This would be due
to that the soliton width is smaller for the stronger Coulomb
repulsions and the portion of regions with nearly perfect
dimerization strengths is larger.  The increase of the ratio
saturates at about 5 percent in Fig. 5(a) and at about 7
percent in Fig. 5(b).  The soliton-continuum exciton becomes
like a free exciton at larger concentrations owing to the
formation of the soliton band.

\section{Bipolaron lattice systems: Confinement effects}

There exist only two kinds of degenerate conjugated polymers.
They are {\sl trans}-polyacetylene and pernigraniline base.
All the other conjugated polymers have nondegenerate ground
states.  Therefore, it is also interesting to look at exciton
effects in this type of polymers.  The structures of nondegenerate
conjugated polymers are generally complex, including aromatic
rings and side chains.  By this reason, the simple SSH-type
models directly simulate only the structures of linear-chain
polymers: {\sl trans}- and {\sl cis}-polyacetylenes.  However,
the SSH model with the non-zero BK term has the nondegenerate
ground state which is one of the general properties described
by the simple SSH-type models.  Therefore, it is of general
interests to look at excitonic effects on nondegenerate
conjugated polymers.  We shall consider the model Eq. (1) with
$\delta_0 = 0.02$.  This small BK term gives rise to the large
energy difference between the ground state and the excited state
with the reversed phase of the bond alternation pattern, after
calculations of adiabatic approximation with the full lattice
relaxation.  The Coulomb and lattice parameters are the same
as in the previous section.

Figure 6 shows the static lattice configuration and the
excess-electron density distribution for $N=100, \nel = 104, U=4t$,
and $V=2t$.  Figure 6(a) displays the lattice configuration.
The ground state with the positive bond variable is more
stable than the state with the negative bond variable.  Therefore,
the region with the negative bond variable becomes smaller
than in Fig. 1(a), and two neighboring solitons come closer each
other to form a bipolaron.  There are two bipolarons in Fig. 6(a).
The electron distribution pattern in Fig. 6(b) reflects the
fact that two solitons are confined to form a bipolaron.  If
the confinement is more stronger, two peaks in the charge density
distribution change into a single peak.

As we have done in the previous section, we shall calculate
optical absorption spectra by using the HF wavefunctions only
as well as by performing single-CI calculations.  Figure 7
shows the results of HF absorption (thin lines) and those of
HF plus single-CI absorption (thick lines).  The following
parameters are the same as in Fig. 6: $U = 2V = 4t$ and
$N = 100$.  The electron number is $\nel = 102$ for Fig. 7(a)
and $\nel = 104$ for Fig. 7(b).  The broadening $\gamma = 0.05t$ is
used. The major difference between the HF and HF-CI absorption
is that the overall feature in the HF absorption shifts to lower
energies in the HF-CI one.  This is one of the exciton effects.

A single bipolaron has two midgap states.  Then, there are
two midgap bands in the bipolaron lattice system.  The number
of excitons is three.  We shall call them as follows: the
photo-excited state from the upper bipolaron band to the conduction
band as the ``lower bipolaron-continuum exciton", the exciton
from the lower bipolaron band to the conduction band as the ``upper
bipolaron-continuum exciton", and the exciton between the continuum
states as the ``intercontinuum exciton".   In Fig. 7(a), the optical
gaps of the lower bipolaron-continuum exciton, the upper
bipolaron-continuum exciton, and the inter-continuum exciton
are about 0.5$t$, 0.7$t$, and 1.6$t$, respectively.  In Fig. 7(b),
they are about 0.6$t$, 0.9$t$, and 1.6$t$, respectively.  The total
oscillator strength of the upper bipolaron-continuum exciton is
always much smaller than that of the other two excitons.  This
property is already seen in the free electron case and is
independent of the magnitudes of $U$ and $V$.  As the
concentration of the bipolarons increases, the oscillator
strength of the lower bipolaron-continuum exciton enhances
rapidly, and that of the intercontinuum exciton becomes smaller.

In order to analyze the concentration dependences systematically,
we perform calculations for several combinations of the system
size $N$ and the electron number $\nel$.  The electron number
is always even, and then the bipolaron number is half of the
excess-electron number $\nel - N$.  Figure 8 summarizes the
optical gaps of three kinds of excitons (single-CI
description), plotted against the excess-electron
concentration: $(\nel - N)/N$.  We take two combinations of
Coulomb parameters: $(U,V) = (2t, 1t)$ [Fig. 8(a)] and $(4t, 2t)$
[Fig. 8(b)].  The optical gaps of the intercontinuum exciton,
and upper (lower) bipolaron-continuum excitons are shown by the
filled, and crossed (open) squares, respectively.  The arrays of
the plots behave smoothly, so it seems that the size effects
are small.  The optical gaps of the intercontinuum exciton and
upper bipolaron-continuum exciton increase gradually as functions
of the excess-electron concentration.  But, the increase of
the lowest optical gap is suppressed, and it is nearly constant
for concentrations larger than about 5 \%.  The similar behaviors
have been seen in Fig. 3 for the soliton lattice case.

Finally, we consider character of optical excitations by
looking at the ratios of the total oscillator strengths
of lower and upper bipolaron-continuum excitons.
Figure 9 shows the results plotted against the excess-electron
concentration.  The squares are the data of the lower
bipolaron-continuum excitons, and circles are the data
of the upper ones.  The open symbols are the data
of the HF absorption, and the closed ones are for the HF-CI
calculations.  The ratios in the HF-CI absorption spectra
are calculated as we have done in the soliton lattice case.
Here, we assign the optical excitation to the exciton whose
component is the largest among three excitons.  The closed squares
have the larger ratio than the open ones, due to exciton effects.
The increase near the zero concentration of the lower
bipolaron-continuum exciton is suppressed for the stronger
$U$ and $V$ of Fig. 9(b) than in Fig. 9(a).  In other words,
the increase of the ratio is steeper for weaker Coulomb
interactions.  This would be due to that the bipolaron
width is smaller for the stronger Coulomb repulsions and
the portion of regions with nearly perfect dimerization
strengths is larger.  The increase of the ratio nearly
saturates at about 5 percent.  The lower bipolaron-continuum
exciton becomes like a free exciton at larger concentrations.
In contrast, the ratio of the upper bipolaron-continuum exciton
is always small, and it is smaller than about 0.20.
This smallness is already seen in the free electron case,
and persists when there are Coulomb interactions.
The same property in the optical spectra of the polaron
and the bipolaron has been discussed in [13].

\section{Summary}

We have looked at exciton effects on soliton and bipolaron
lattice states in a model of the interacting electron lattice
system with long-range Coulomb interactions.  The Hartree-Fock
approximation and the single-CI method have been used to obtain
optical absorption spectra.  We have discussed the following
properties:\\
(1)  By comparison of the HF absorption with the HF-CI one,
we have seen exciton effects which are similar to those,
discussed for the half-filled systems in refs. [5,7].  The
attraction between the excited electron and the remaining
hole makes the excitation energy smaller when the correlations
are taken into account by the single-CI.  The oscillator strengths
of the lower excited states become relatively larger than
in the HF calculations.  \\
(2)  We have looked at variations of relative oscillator
strengths of two or three kinds of excitons described by
the single-CI.  While the excess-electron concentration is
small, the ratio of the oscillator strengths of the lowest
exciton increases almost linearly.  And, more than 80 percent
of the oscillator strengths accumulate at the lowest exciton
when the excess-electron concentration is larger than about
5 percent.  It seems that this accumulation is very rapid
as the concentration increases. \\
(3)  The variations of the optical gaps and the ratios of the
oscillator strengths as functions of the concentration of the
nonlinear excitations (solitons or bipolarons) are rather smooth.
It seems that how to systematically deal with such kinds of
the numerical data obtained from numerical diagonalizations
of finite systems is not fully investigated yet.  We have
looked variations as functions of the concentration, and
have found the smooth variations.  This fact was not stressed
upon earlier, and plotting as a function of the concentration
will be one of the useful methods for future investigations
of systems with nonlinear excitations.

\mbox{}

\noindent
{\bf Acknowledgements}\\
The author acknowledges useful discussion with Prof. S. Stafstr\"{o}m
and Dr. Akira Takahashi.

\pagebreak
\begin{flushleft}
{\bf References}
\end{flushleft}

\noindent
$[1]$ Kuroda S and Shirakawa H 1987 {\sl Phys. Rev.} B {\bf 35} 9380\\
$[2]$ Su W P, Schrieffer J R and Heeger A J 1980
{\sl Phys. Rev.} B {\bf 22} 2099\\
$[3]$ Yonemitsu K and Wada Y 1988 {\sl J. Phys. Soc. Jpn.}
{\bf 57} 3875; Sasai M and Fukutome H 1983 {\sl Prog. Theor. Phys.}
{\bf 69} 373\\
$[4]$ Long S M, Sun Y, MacDiarmid A G and Epstein A J
1994 {\sl Phys. Rev. Lett.} {\bf 72} 3210\\
$[5]$ Abe S, Yu J and Su W P 1992 {\sl Phys. Rev.} B {\bf 45} 8264\\
$[6]$ Guo D, Mazumdar S, Dixit S N, Kajzar F,
Jarka F, Kawabe Y and Peyghambarian N 1993 {\sl Phys. Rev.}
B {\bf 48} 1433\\
$[7]$ Takahashi A and Mukamel S 1994 {\sl J. Chem. Phys.}
{\bf 100} 2366\\
$[8]$ Abe S, Schreiber M, Su W P and Yu J 1992
{\sl Phys. Rev.} B {\bf 45} 9432\\
$[9]$ Horovitz B 1981 {\sl Phys. Rev. Lett.} {\bf 46} 742\\
$[10]$ Brazovskii S A and Kirova N N 1981
{\sl JETP Lett.} {\bf 33} 4\\
$[11]$ Terai A and Ono Y 1986 {\sl J. Phys. Soc. Jpn.}
{\bf 55} 213\\
$[12]$ Abe S, Schreiber M, Su W P and Yu J 1992
{\sl Mol. Cryst. Liq. Cryst.} {\bf 217} 1.\\
$[13]$ Shimoi Y, Abe S and Harigaya K
{\sl Mol. Cryst. Liq. Cryst.} (to be published).\\

\pagebreak

\begin{flushleft}
{\bf FIGURE CAPTIONS}
\end{flushleft}

\mbox{}

\noindent
Fig. 1. (a)  The dimerization order parameter,
$(-1)^n (y_{n+1} - y_n)/2$, and (b) the excess-electron
density, $(\rho_{n-1} + 2 \rho_n + \rho_{n+1})/4$.
The parameters are $\delta = 0$, $U=4t$, $V=2t$, $N=100$,
and $\nel = 104$.  See the text for the other parameters.

\mbox{}

\noindent
Fig. 2.  The optical absorption spectra calculated with the HF
wavefunctions for (a) $(N,\nel) = (101,102)$, (b) (100,102),
and (c) (100,104).  The broadening $\gamma = 0.05t$ is used.
The units of the abscissa are arbitrary.  The component of the
electron-hole excitation between the soliton and conduction
bands are shown by dots which are connected with thin lines.
The energy positions of the optical gaps are shown by the triangles
at the top of each figure.

\mbox{}

\noindent
Fig. 3.  The optical absorption spectra calculated with the HF plus
single-CI wavefunctions for (a) $(N,\nel) = (101,102)$, (b) (100,102),
and (c) (100,104).  The broadening $\gamma = 0.05t$ is used.
The units of the abscissa are arbitrary.  The component of the
electron-hole excitation between the soliton and conduction
bands are shown by dots which are connected with thin lines.
The energy positions of the optical gaps are shown by the triangles
at the top of each figure.

\mbox{}

\noindent
Fig. 4.  The optical gaps in the single-CI of the ``intercontinuum
exciton" (filled squares) and of the ``soliton-continuum exciton"
(open squares), plotted against the soliton concentration.
Coulomb interaction parameters are $(U,V) = (2t,1t)$ for (a),
and $(4t,2t)$ for (b).

\mbox{}

\noindent
Fig. 5.  The ratio of the total oscillator strength of the
"soliton-continuum exciton" as a function of the soliton concentration.
The open squares are the data for the HF absorption, while the filled
ones for the HF-CI absorption.  Coulomb interaction parameters are
$(U,V) = (2t,1t)$ for (a), and $(4t,2t)$ for (b).

\mbox{}

\noindent
Fig. 6. (a)  The dimerization order parameter,
$(-1)^n (y_{n+1} - y_n)/2$, and (b) the excess-electron
density, $(\rho_{n-1} + 2 \rho_n + \rho_{n+1})/4$.
The parameters are $\delta = 0.02$, $U=4t$, $V=2t$, $N=100$,
and $\nel = 104$.  See the text for the other parameters.

\mbox{}

\noindent
Fig. 7.  The optical absorption spectra calculated with the HF
(thin lines) and HF-CI (full lines) wavefunctions
for (a) $(N,\nel) = (100,102)$ and (b) (100,104).
The broadening $\gamma = 0.05t$ is used.
The units of the abscissa are arbitrary.

\mbox{}

\noindent
Fig. 8.  The optical gaps in the single-CI of the ``intercontinuum
exciton" (filled squares), of the ``upper bipolaron-continuum exciton"
(crossed squares), and of the ``lower bipolaron-continuum exciton"
(open squares), plotted against the excess-electron concentration.
Coulomb interaction parameters are $(U,V) = (2t,1t)$ for (a),
and $(4t,2t)$ for (b).

\mbox{}

\noindent
Fig. 9.  The ratio of the total oscillator strength of the
"upper (lower) bipolaron-continuum exciton" as functions
of the excess-electron concentration.  The squares are for
the lower exciton, and the circles are for the upper exciton.
The open symbols are the data for the HF absorption, while the filled
ones for the HF-CI absorption.  Coulomb interaction parameters are
$(U,V) = (2t,1t)$ for (a), and $(4t,2t)$ for (b).

\end{document}